\documentclass{elsart}

\usepackage{adnd}
\usepackage{longtable}

\usepackage{mathptmx}

\usepackage{amsmath}

\usepackage{amssymb,amstext}
\usepackage[pdftex,
 letterpaper=true,
 hyperindex=true,
 breaklinks=true,
 colorlinks=false,
 citecolor=blue,
 pdftitle={},
 pdfauthor={}]
{hyperref}

\usepackage{graphicx}

\begin{document}

\begin{frontmatter}

\journal{Atomic Data and Nuclear Data Tables}

\copyrightholder{Elsevier Science}

\runtitle{Iron}
\runauthor{Schuh}


\title{Discovery of the Iron Isotopes}


\author{A.~Schuh},
\author{A.~Fritsch},
\author{M.~Heim},
\author{A.~Shore},
\and
\author{M.~Thoennessen\corauthref{cor}}\corauth[cor]{Corresponding author.}\ead{thoennessen@nscl.msu.edu}

\address{National Superconducting Cyclotron Laboratory and \\ Department of Physics and Astronomy, Michigan State University, \\East Lansing, MI 48824, USA}

\date{July 7, 2009} 

\begin{abstract}
Twenty-eight iron isotopes have so far been observed; the discovery of these isotopes is discussed. For each isotope a brief summary of the first refereed publication, including the production and identification method, is presented.
\end{abstract}

\end{frontmatter}





\newpage
\tableofcontents
\listofDtables

\vskip5pc

\section{Introduction}\label{s:intro}

In this seventh paper in the series of the discovery of isotopes, the discovery of the iron isotopes is discussed. Previously, the discovery of cerium \cite{Gin09}, arsenic \cite{Sho09}, gold \cite{Sch09}, tungsten \cite{Fri09}, krypton \cite{Hei09} and einsteinium \cite{Bur09} isotopes was discussed. The purpose of this series is to document and summarize the discovery of the isotopes. Guidelines for assigning credit for discovery are (1) clear identification, either through decay-curves and relationships to other known isotopes, particle or $\gamma$-ray spectra, or unique mass and Z-identification, and (2) publication of the discovery in a refereed journal. The authors and year of the first publication, the laboratory where the isotopes were produced as well as the production and identification methods are discussed. When appropriate, references to conference proceedings, internal reports, and theses are included. When a discovery includes a half-life measurement, we compared the measured value to the currently adapted value taken from the NUBASE evaluation \cite{Aud03} which is based on ENSDF database \cite{ENS08}. In cases where the reported half-life differed significantly from the adapted half-life (up to approximately a factor of two), we searched the subsequent literature for indications that the measurement was erroneous. If that was not the case we credited the authors with the discovery in spite of the inaccurate half-life.

\section{Discovery of $^{45-72}$Fe}

Twenty-eight iron isotopes from A = $45-72$ have been discovered so far; these include 4 stable, 10 proton-rich and 14 neutron-rich isotopes. According to the HFB-14 model \cite{Gor07} iron isotopes are predicted to be stable with respect to one neutron emission out to $^{81}$Fe for the odd-mass isotopes and two-neutron emission out to $^{90}$Fe for the even-mass isotopes. At the proton dripline one more isotope, $^{44}$Fe, is predicted to be stable with respect to nucleon emission. Thus, there remain 15 isotopes to be discovered. No additional nuclei beyond the proton dripline are predicted to live long enough to be observed \cite{Tho04}. Over 60\% of all possible iron isotopes have been produced and identified so far.

\begin{figure}
	\centering
	\includegraphics[scale=.5]{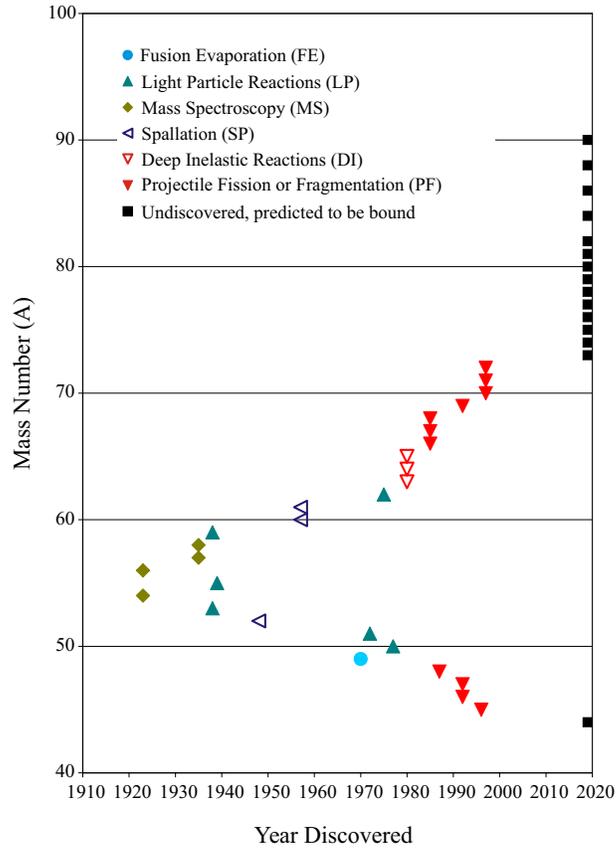}
	\caption{Iron isotopes as a function of time they were discovered. The different production methods are indicated. The solid black squares on the right hand side of the plot are isotopes predicted to be bound by the HFB-14 model.}
\label{f:year}
\end{figure}

Figure \ref{f:year} summarizes the year of first discovery for all iron isotopes identified by the method of discovery. The range of isotopes predicted to exist is indicated on the right side of the figure. The radioactive iron isotopes were produced using heavy-ion fusion evaporation (FE), light-particle reactions (LP), spallation reactions (SP), deep-inelastic reactions (DI), and projectile fragmentation or fission (PF). The stable isotopes were identified using mass spectroscopy (MS). Heavy ions are all nuclei with an atomic mass larger than A = 4 \cite{Gru77}. Light particles also include neutrons produced by accelerators. In the following paragraphs, the discovery of each iron isotope is discussed in detail.

\subsection*{$^{45}$Fe}\vspace{-0.85cm}

In their paper \textit{First Observation of the T$_{z}=-$7/2 Nuclei $^{45}$Fe and $^{49}$Ni}, Blank \textit{et al.} reported the discovery of $^{45}$Fe in 1996 at the Gesellschaft f\"{u}r Schwerionenforschung (GSI) in Germany \cite{Bla96}. A 600 A$\cdot$MeV $^{58}$Ni beam bombarded a 4~g/cm$^2$ thick beryllium target and isotopes were separated with the projectile-fragment separator FRS. $^{45}$Fe was identified by time-of-flight, $\Delta$E, and B$\rho$ analysis. ``In the entire Z versus A/Z plot ... only one background event can be identified... This high background suppression enables us to conclude even with only three and five counts on the first observation of $^{45}$Fe and $^{49}$Ni, respectively.'' The half-life was estimated to be greater than 350~ns based on the flight time through the FRS. In 1992 the non-observation of $^{45}$Fe in a projectile fragmentation experiment had led to the suggestion that $^{45}$Fe was probably not stable with respect to particle emission \cite{Bor92}.

\subsection*{$^{46,47}$Fe}\vspace{-0.85cm}

$^{46}$Fe and $^{47}$Fe were discovered by Borrel \textit{et al.} at Grand Accelerateur National D'ions Lourds (GANIL) in France in 1992, as reported in the paper \textit{The decay modes of proton drip-line nuclei with A between 42 and 47} \cite{Bor92}. A 69 A$\cdot$MeV $^{58}$Ni beam was incident on a natural nickel target and the projectile fragments were separated using the Ligne d'Ions Super Epluch\'{e}s (LISE) spectrometer. The isotopes were identified by time of flight and energy loss measurements. ``A three hour run leads to the first identification of $^{47}$Fe with 23 counts'' and ``another step is taken towards the proton dripline with the identification of $^{46}$Fe. Sixteen events are obtained in nineteen hours.'' The half-life of $^{46}$Fe was experimentally determined via maximum-likelihood analysis of the time spectrum to be 20$^{+20}_{-8}$ms; this value agrees with the presently accepted value of 9(4)~ms. The half-life of $^{47}$Fe was also determined via maximum-likelihood analysis of the time spectrum to be 27$^{+32}_{-10}$ms. Pougheon \textit{et al.} had observed one count of $^{47}$Fe at GANIL in 1987, but attributed the uncertain event to background \cite{Pou87}.

\subsection*{$^{48}$Fe}\vspace{-0.85cm}

The 1987 paper \textit{Direct Observation of New Proton Rich Nuclei in the Region 23$\leq$Z$\leq$29 Using A 55A.MeV $^{58}$Ni Beam}, reported the first observation of $^{48}$Fe at GANIL by Pougheon \textit{et al.} \cite{Pou87}. The fragmentation of a 55 A$\cdot$MeV $^{58}$Ni beam on a nickel target was used to produce proton-rich isotopes which were separated with the LISE spectrometer. Energy loss, time of flight, and magnetic rigidity measurements were made such that ``$^{48}$Fe is clearly identified with 27 counts.''

\subsection*{$^{49}$Fe}\vspace{-0.85cm}

$^{49}$Fe was first observed by Cerny \textit{et al.} in 1970 and reported in the paper $^{49}$\textit{Fe A New T${_z}=-$3/2 Delayed-Proton Emitter} \cite{Cer70}. The reaction $^{40}$Ca($^{12}$C,3n) using 65 A$\cdot$MeV carbon ions accelerated by the Harwell variable-energy cyclotron was used to produce $^{49}$Fe. Beta-delayed protons were measured with a semiconductor telescope consisting of two surface-barrier detectors. ``Figure 1(b) presents a proton spectrum from $^{49}$Fe produced from a 2.2 mg/cm$^2$ Ca target. A single peak corresponding to a c.m. energy of 1.96(0.5) MeV, after correction for energy loss in the target, dominates the proton decay.'' The half-life was measured to be 75(10)~ms, which is consistent with the accepted value of 70(3)~ms.

\subsection*{$^{50}$Fe}\vspace{-0.85cm}

In the paper \textit{Mass measurements of the proton-rich nuclei $^{50}$Fe and $^{54}$Ni}, Tribble \textit{et al.} reported the discovery of $^{50}$Fe in 1977 \cite{Tri77}. Alpha particles accelerated to 110 MeV with the Texas A\&M University 88-inch Cyclotron were used to produce the reaction $^{54}$Fe($^{4}$He,$^{8}$He) and the ejectiles were observed at the focal plane of an Enge split-pole magnetic spectrograph. ``The experiments provide the first observation and subsequent mass measurement of the proton-rich nuclei $^{50}$Fe and $^{54}$Ni.'' The measured $\beta$-decay energy was 7.12(6)~MeV which was used to estimate a half-life of 200~ms; this is close to the adapted value of 155(11)~ms.

\subsection*{$^{51}$Fe}\vspace{-0.85cm}

In a paper entitled \textit{New Proton-Rich Nuclei in the f${_{7/2}}$ Shell}, Proctor \textit{et al.} described the discovery of $^{51}$Fe in 1972 \cite{Pro72}. The Michigan State University sector-focused cyclotron accelerated $^{3}$He to 70.8~MeV and the reaction $^{54}$Fe($^{3}$He,$^{6}$He) was used to produce $^{51}$Fe. The outgoing $^6$He particles were detected in the focal plane of an Enge split-pole magnetic spectrograph. ``The $^{51}$Fe ground state (J$^\pi$ = 5/2$^-$) is even more weakly populated, but is unambiguously identified in a number of spectra.''

\subsection*{$^{52}$Fe}\vspace{-0.85cm}

\textit{Products of High Energy Deuteron and  Helium Ion Bombardments of Copper} presented the first observation of $^{52}$Fe by Miller \textit{et al.} in 1948 \cite{Mil48}. The bombardment of natural copper with 190 MeV deuterons from the Berkeley 184-inch frequency-modulated cyclotron was used to produce $^{52}$Fe in a spallation type reaction. ``An aluminum absorption curve of the parent-daughter equilibrium mixture showed, in addition to a component of \textit{ca.} 2.3 Mev attributable to the 21-min $^{52}$Mn, a component of \textit{ca.} 0.55-Mev maximum energy presumably as a result of the 7.8-hour parent, assigned to $^{52}$Fe.'' This measured half-life agrees with the presently accepted value of 8.275(8)~h.

\subsection*{$^{53}$Fe}\vspace{-0.85cm}

In \textit{Radioactive Isotopes of Iron}, Livingood and Seaborg reported the production of $^{53}$Fe in 1938 \cite{Liv38}. The isotope was produced in the reaction $^{50}$Cr($\alpha$,n)$^{53}$Fe with 16 MeV $\alpha$-particles accelerated by the Berkeley cyclotron. The decay curves of the produced radioactivity were measured with a quartz fiber electroscope following chemical separation. The authors ``believe the 9-minute activity to be due to Fe$^{53}$ rather than to Fe$^{55}$ because: (1) it is not produced by deuteron or slow neutron bombardment of Fe, (2) it is produced by fast neutrons on Fe, (3) attempts to produce $^{55}$Fe by other reactions have not disclosed a 9 minute activity.'' The half-life was determined to be 8.9(2)~m which is close to the accepted value of 8.51(2)~m. In 1937, Ridenour and Henderson had observed a 9-minute activity; however, they were unable to make the unique mass assignment and attributed it to either the reaction $^{50}$Cr($\alpha$,n)$^{53}$Fe or the reaction $^{52}$Cr($\alpha$,n)$^{55}$Fe \cite{Rid37a}. In an even earlier publication, they had preferred the later assignment \cite{Rid37b}.

\subsection*{$^{54}$Fe}\vspace{-0.85cm}

In his 1923 paper \textit{The Mass Spectra of Elements - Part IV}, Aston mentioned the possible first observation of $^{54}$Fe \cite{Ast23}. Using his mass spectrograph in Cavendish he investigated iron with the volatile carbonyl. ``The faint line at 54 may possibly be an isotope, but this is by no means certain.'' Aston confirmed the observation in 1925 \cite{Ast25}.

\subsection*{$^{55}$Fe}\vspace{-0.85cm}

Livingood and Seaborg observed $^{55}$Fe in 1938 described in the paper \textit{Long-Lived Radioactive Fe$^{55}$} \cite{Liv39}. Iron samples bombarded with deuterons from the Berkeley cyclotron described in a previous publication \cite{Liv38} were measured for a period of 22 months. ``These facts assure us that Fe$^{55}$ has been formed through Fe${54}$(d,p)Fe$^{55}$ with the activity probably leading to stable Mn$^{55}$ either by positron emission or by K-electron capture.'' The counting time was not sufficient to extract a reliable half-life measurement and only a lower limit of one year was determined. The currently accepted half-life value is 2.737(10)~y.

\subsection*{$^{56}$Fe}\vspace{-0.85cm}

$^{56}$Fe was first identified at Cavendish in 1923 by Aston \cite{Ast23} as reported in \textit{The Mass Spectra of Elements - Part IV}. Volatile iron carbonyl was used to obtain the mass spectrum. ``The only line which can be ascribed with certainty to iron is the one at 56. Thirteen independent measurements of the principal line relative to other lines on the plate gave values of its mass which were very consistent and had a mean of 55.94.''

\subsection*{$^{57}$Fe}\vspace{-0.85cm}

In 1935 Aston discovered $^{57}$Fe at Cavendish and described the results in his article \textit{The Isotopic Constitution and Atomic Weights of Hafnium, Thorium, Rhodium, Titanium, Zirconium, Calcium, Gallium, Silver, Carbon, Nickel, Cadmium, Iron and Indium} \cite{Ast35}. Aston used a pure sample of the carbonyl in the spectrograph. ``In addition to the strong isotope 56 and a weak one, 54, previously known, a third, 57 was revealed.''

\subsection*{$^{58}$Fe}\vspace{-0.85cm}

 The existence of $^{58}$Fe was demonstrated by deGier and Zeeman at the University of Amsterdam in 1935 and reported in the paper \textit{The Isotopic Constitution of Iron} \cite{deG35}. DeGier and Zeeman succeeded with the identification of $^{58}$Fe with a very pure sample of carbonyl. ``With properly chosen canals the intensity of the iron lines could be increased so far that isotope 58 can be seen in the reproduction... The appearance of line 58 could now be followed closely when varying the circumstances of the experiments. In this way we obtained several convincing plates of the new isotope.'' In early 1935 Aston was not confident in the observation of $^{58}$Fe: ``Line 58 was present but weakened as the work proceeded and was most probably due to traces of nickel still left in the tube'' \cite{Ast35}.

\subsection*{$^{59}$Fe}\vspace{-0.85cm}

$^{59}$Fe was discovered by Livingood and Seaborg in 1938 as reported in \textit{Radioactive Isotopes of Iron} \cite{Liv38}. $^{59}$Fe was produced in the reactions $^{58}$Fe(d,p) and $^{59}$Co(n,p) with 5.5 MeV deuterons from the Berkeley cyclotron. The neutron irradiation was performed by placing the target next to the cyclotron during the bombardment of deuterons on lithium. The decay curves of the produced radioactivity were measured with a quartz fiber electroscope following chemical separation. ``It is at once apparent that only Fe$^{59}$ can be negative electron active. Furthermore, the only radio-iron that can be made from cobalt with neutron is Fe$^{59}$, so that we are justified in ascribing the 47-day activity to this isotope.'' The measured half-life of 47(3)~d is consistent with the accepted half-life of 44.495(9)~d.
Livingood {\it et al.} had reported a 40~d iron activity in 1937 without attributing it to a specific isotope \cite{Liv37}.

\subsection*{$^{60}$Fe}\vspace{-0.85cm}

The discovery of $^{60}$Fe was described by Roy and Kohman in the 1957 paper \textit{Iron 60} \cite{Roy57}. A copper target was bombarded with 400~MeV protons from the Carnegie synchocyclotron in Pittsburgh and $^{60}$Fe was produced in a spallation reaction. The mass assignment was made through the observation of the decay to the $^{60m}$Co daughter following chemical separation. ``From this, the activity ratio of Fe$^{60}$ and Fe$^{59}$, 45 days, the half-life of Fe$^{60}$ can be derived. The result is $\sim 3\cdot 10^5$ years, uncertain by a factor of 3 because of the approximate nature of the measurements and calculations.'' This half-life is somewhat smaller than the accepted value of 1.5(3)$\cdot$10$^{6}$~y.

\subsection*{$^{61}$Fe}\vspace{-0.85cm}

Ricci \textit{et al.} were the first to produce $^{61}$Fe in 1957 and published the results in the article \textit{A New Isotope of Iron $^{61}$Fe} \cite{Ric57}. $^{61}$Fe was produced in the spallation of nickel and copper targets in Buenos Aires, Argentina. ``Mass number 61 was assigned to the new iron isotope because it decays to 99 minutes $^{61}$Co, already known.'' The half-life was measured to be 6.0(5)~m. This is consistent with the accepted value of  5.98(6)~m.

\subsection*{$^{62}$Fe}\vspace{-0.85cm}

In the 1975 paper \textit{Decay of the New Isotope $^{62}$Fe}, Franz \textit{et al.} reported the first observation of $^{62}$Fe \cite{Fra75}. Neutrons between 25 and 200~MeV generated by 200~MeV protons from the Brookhaven AGS linac injector bombarded a nickel oxide target enriched to 96\% $^{64}$Ni. $^{62}$Fe was produced with the $^{64}$Ni(n,2pn) reaction. Gamma spectra were measured following chemical separation. ``The mass assignment must be to $^{62}$Fe because the appropriate growth and decay were observed of 1.5-m $^{62}$Co in the chemically purified iron sample.'' The half-life of 68(2)~s is currently the only measured value for $^{62}$Fe.

\subsection*{$^{63-65}$Fe}\vspace{-0.85cm}

Guerreau \textit{et al.} reported the discovery of $^{63}$Fe, $^{64}$Fe and $^{65}$Fe in the 1980 paper \textit{Seven New Neutron Rich Nuclides Observed in Deep Inelastic Collisions of 340 MeV $^{40}$Ar on $^{238}$U} \cite{Gue80}. A 340 MeV $^{40}$Ar beam accelerated by the Orsay ALICE accelerator facility bombarded a 1.2 mg/cm$^2$ thick UF$_4$ target supported by an aluminum foil. The isotopes were identified using two $\Delta$E-E telescopes and two time of flight measurements. ``The new nuclides $^{54}$Ti, $^{56}$V, $^{58-59}$Cr, $^{61}$Mn, $^{63-64}$Fe, have been produced through $^{40}$Ar + $^{238}$U reactions.'' At least twenty counts were recorded for these isotopes. The tentative observation of $^{65}$Fe was mentioned. An inspection of the spectrum indicates at least 6 events of $^{65}$Fe. Breuer \textit{et al.} detected $^{63}$Fe independently only a few months later \cite{Bre80}.

\subsection*{$^{66-68}$Fe}\vspace{-0.85cm}

The 1985 paper \textit{Production and Identification of New Neutron-Rich Fragments from 33 MeV/u $^{86}$Kr Beam in the 18$\leq$Z$\leq$27 Region} by Guillemaud-Mueller \textit{et al.} reported the first observation of $^{66}$Fe, $^{67}$ Fe and $^{68}$Fe \cite{Gui85}. The 33 MeV/u $^{86}$Kr beam bombarded tantalum targets and the  fragments were separated with the GANIL triple-focusing analyser LISE. ``Each particle is identified by an event-by-event analysis. The mass A is determined from the total energy and the time of flight, and Z by the $\Delta$E and E measurements.''

\subsection*{$^{69}$Fe}\vspace{-0.85cm}

In their paper \textit{New neutron-rich isotopes in the scandium-to-nickel region, produced by fragmentation of a 500 MeV/u $^{86}$Kr beam}, Weber \textit{et al.} presented the first observation of $^{69}$Fe in 1992 at GSI \cite{Web92}. $^{69}$Fe was produced in the fragmentation reaction of a 500 A$\cdot$MeV $^{86}$Kr beam from the heavy-ion synchroton SIS on a beryllium target and separated with the zero-degree spectrometer FRS. ``The isotope identification was based on combining the values of B$\rho$, time of flight (TOF), and energy loss ($\triangle$E) that were measured for each ion passing through the FRS and its associated detector array.'' Twelve counts of $^{69}$Fe were recorded.

\subsection*{$^{70-72}$Fe}\vspace{-0.85cm}

Bernas \textit{et al.} observed $^{70}$Fe, $^{71}$Fe and $^{72}$Fe for the first time in 1997 as reported in their paper \textit{Discovery and cross-section measurement of 58 new fission products in projectile-fission of 750$\cdot$A MeV $^{238}$U} \cite{Ber97}. Uranium ions were accelerated to 750 A$\cdot$MeV by the GSI UNILAC/SIS accelerator facility and bombarded a beryllium target. The isotopes produced in the projectile-fission reaction were separated using the fragment separator FRS and the nuclear charge Z for each was determined by the energy loss measurement in an ionization chamber. ``The mass identification was carried out by measuring the time of flight (TOF) and the magnetic rigidity B$\rho$ with an accuracy of 10$^{-4}$.'' Two hundred counts of $^{70}$Fe, 39 counts of $^{71}$Fe, and two counts of $^{72}$Fe were observed.

\section{Summary}
The discovery of the iron isotopes has been mostly uncontroversial. The activities of only two isotopes ($^{53}$Fe and $^{59}$Fe) were detected before they could be assigned to the specific isotope. Prior to the discovery of $^{45}$Fe it was claimed to be potentially unstable due to the non-observation in a fragmentation experiment.

\ack

This work was supported by the National Science Foundation under grants No. PHY06-06007 (NSCL) and PHY07-54541 (REU). MH was supported by NSF grant PHY05-55445.


\newpage

\section*{EXPLANATION OF TABLE}\label{sec.eot}
\addcontentsline{toc}{section}{EXPLANATION OF TABLE}

\renewcommand{\arraystretch}{1.0}

\begin{tabular*}{0.95\textwidth}{@{}@{\extracolsep{\fill}}lp{5.5in}@{}}
\textbf{TABLE I.}
	& \textbf{Discovery of Iron Isotopes }\\
\\

Isotope & Iron isotope \\
Author & First author of refereed publication \\
Journal & Journal of publication \\
Ref. & Reference \\
Method & Production method used in the discovery: \\
 & FE: fusion evaporation \\
 & LP: light-particle reactions (including neutrons) \\
 & MS: mass spectroscopy \\
 & DI: deep-inelastic reactions \\
 & SP: spallation \\
 & PF: projectile fragmentation or projectile fission \\
Laboratory & Laboratory where the experiment was performed\\
Country & Country of laboratory\\
Year & Year of discovery \\
\end{tabular*}
\label{tableI}

\newpage
\datatables

\setlength{\LTleft}{0pt}
\setlength{\LTright}{0pt}


\setlength{\tabcolsep}{0.5\tabcolsep}

\renewcommand{\arraystretch}{1.0}


\begin{longtable}[c]{%
@{}@{\extracolsep{\fill}}r@{\hspace{5\tabcolsep}} llllllll@{}}
\caption[Discovery of Iron Isotopes]%
{Discovery of Iron isotopes}\\[0pt]
\caption*{\small{See page \pageref{tableI} for Explanation of Tables}}\\
\hline
\\[100pt]
\multicolumn{8}{c}{\textit{This space intentionally left blank}}\\
\endfirsthead
Isotope & Author & Journal & Ref. & Method & Laboratory & Country & Year \\
$^{45}$Fe & B. Blank & Phys. Rev. Lett. & Bla96 & PF & Darmstadt & Germany &1996 \\
$^{46}$Fe & V. Borrel & Z. Phys. A & Bor92 & PF & GANIL & France &1992 \\
$^{47}$Fe & V. Borrel & Z. Phys. A & Bor92 & PF & GANIL & France &1992 \\
$^{48}$Fe & F. Pougheon & Z. Phys. A & Pou87 & PF & GANIL & France &1987 \\
$^{49}$Fe & J. Cerny & Phys. Rev. Lett. & Cer70 & FE & Harwell & UK &1970 \\
$^{50}$Fe & R.E. Tribble & Phys. Rev. C & Tri77 & LP & Texas A\&M & USA &1977 \\
$^{51}$Fe & I.D. Proctor & Phys. Rev. Lett. & Pro72 & LP & Michigan State & USA &1972 \\
$^{52}$Fe & D.R. Miller & Phys. Rev. & Mil48 & SP & Berkeley & USA &1948 \\
$^{53}$Fe & J.J. Livingood & Phys. Rev. & Liv38 & LP & Berkeley & USA &1938 \\
$^{54}$Fe & F.W. Aston & Phil. Mag. & Ast23 & MS & Cavendish & UK &1923 \\
$^{55}$Fe & J.J. Livingood & Phys. Rev. & Liv39 & LP & Berkeley & USA &1939 \\
$^{56}$Fe & F. W. Aston & Phil. Mag. & Ast23 & MS & Cavendish & UK &1923 \\
$^{57}$Fe & F. W. Aston & Proc. Roy. Soc. & Ast35 & MS & Cavendish & UK &1935 \\
$^{58}$Fe & J. deGier & Proc. Akad. Soc. & deG35 & MS & Amsterdam & Netherlands &1935 \\
$^{59}$Fe & J.J. Livingood & Phys. Rev. & Liv38 & LP & Berkeley & USA &1938 \\
$^{60}$Fe & J.-C. Roy & Can. J. Phys. & Roy57 & SP & Pittsburgh & USA &1957 \\
$^{61}$Fe & E. Ricci & Com. Nacl. Ener. Atom. & Ric57 & SP & Buenos Aires & Argentina &1957 \\
$^{62}$Fe & E.-M. Franz & Phys. Rev. C & Fra75 & LP & Brookhaven & USA &1975 \\
$^{63}$Fe & D. Guerreau & Z. Phys. A & Gue80 & DI & Orsay & France &1980 \\
$^{64}$Fe & D. Guerreau & Z. Phys. A & Gue80 & DI & Orsay & France &1980 \\
$^{65}$Fe & D. Guerreau & Z. Phys. A & Gue80 & DI & Orsay & France &1980 \\
$^{66}$Fe & D. Guillemaud-Mueller & Z. Phys. A & Gui85 & PF & GANIL & France &1985 \\
$^{67}$Fe & D. Guillemaud-Mueller & Z. Phys. A & Gui85 & PF & GANIL & France &1985 \\
$^{68}$Fe & D. Guillemaud-Mueller & Z. Phys. A & Gui85 & PF & GANIL & France &1985 \\
$^{69}$Fe & M. Weber & Z. Phys. A & Web92 & PF & Darmstadt & Germany &1992 \\
$^{70}$Fe & M. Bernas & Phys. Lett. B & Ber97 & PF & Darmstadt & Germany &1997 \\
$^{71}$Fe & M. Bernas & Phys. Lett. B & Ber97 & PF & Darmstadt & Germany &1997 \\
$^{72}$Fe & M. Bernas & Phys. Lett. B & Ber97 & PF & Darmstadt & Germany &1997 \\

\end{longtable}

\newpage


\normalsize

\begin{theDTbibliography}{1956He83}

\bibitem[Ast23]{Ast23t} F.W. Aston, Phil. Mag. {\bf 45}, 934 (1923)
\bibitem[Ast35]{Ast35t} F.W. Aston, Proc. Roy. Soc. A {\bf 149}, 396 (1935)
\bibitem[Ber97]{Ber97t} M. Bernas, C. Engelmann, P. Armbruster, S. Czajkowski, F. Ameil, C. B\"ockstiegel, Ph. Dessagne, C. Donzaud, H. Geissel, A. Heinz, Z. Janas, C. Kozhuharov, Ch. Mieh\'e, G. M\"unzenberg, M. Pf\"utzner, W. Schwab, C. St\'ephan, K. S\"ummerer, L. Tassan-Got, and B. Voss, Phys. Lett. B {\bf 415}, 111 (1997)
\bibitem[Bla96]{Bla96t} B. Blank, S. Czajkowski, F. Davi, R. Del Moral, J.P. Dufour, A. Fleury, C. Marchand, M.S. Pravikoff, J. Benlliure, F. Bou\'e, R. Collatz, A. Heinz, M. Hellstr\"om, Z. Hu, E. Roeckl, M. Shibata, K. S\"ummerer, Z. Janas, M. Karny, M. Pf\"utzner, and M. Lewitowicz, Phys. Rev. Lett. {\bf 77}, 2893 (1996)
\bibitem[Bor92]{Bor92t} V. Borrel, R. Anne, D. Bazin, C. Borcea, G.G. Chubarian, R. Del Moral, C. Detraz, S. Dogny, J.P. Dufour, L. Faux, A. Fleury, L.K. Fifield, D. Guillemaud-Mueller, F. Hubert, E. Kashy, M. Lewitowicz, C. Marchand, A.C. Mueller, F. Pougheon, M.S. Pravikoff, M.G. Saint-Laurent, and O. Sorlin, Z. Phys. A {\bf 344}, 135 (1992)
\bibitem[Cer70]{Cer70t} J. Cerny, C.U. Cardinal, H.C. Evans, K.P. Jackson, and N.A. Jelley, Phys. Rev. Lett. {\bf 24}, 1128 (1970)
\bibitem[deG35]{deG35t} J. deGier and P. Zeeman, Proc. Akad. Soc. Amsterdam {\bf 38}, 959 (1935)
\bibitem[Fra75]{Fra75t} E.-M. Franz, S. Katcoff, H.A. Smith, Jr., and T.E. Ward, Phys. Rev. C {\bf 12}, 616 (1975)
\bibitem[Gue80]{Gue80t} D. Guerreau, J. Galin, B. Gatty, and X. Tarrago, Z. Phys. A {\bf 295}, 105 (1980)
\bibitem[Gui85]{Gui85t} D. Guillemaud-Mueller, A.C. Mueller, D. Guerreau, F. Pougheon, R. Anne, M. Bernas, J. Galin, J.C. Jacmart, M. Langevin, F. Naulin, E. Quiniou, and C. Detraz, Z. Phys. A {\bf 322}, 415 (1985)
\bibitem[Liv38]{Liv38t} J.J. Livingood and G.T. Seaborg, Phys. Rev. {\bf 54}, 51 (1938)
\bibitem[Liv39]{Liv39t} J.J. Livingood and G.T. Seaborg, Phys. Rev. {\bf 55}, 1268 (1939)
\bibitem[Mil48]{Mil48t} D.R. Miller, R.C. Thompson, and B.B. Cunningham, Phys. Rev. {\bf 74}, 347 (1948)
\bibitem[Pou87]{Pou87t} F. Pougheon, J.C. Jacmart, E. Quiniou, R. Anne, D. Bazin, V. Borrel, J. Galin, D. Guerreau, D. Guillemaud-Mueller, A.C. Mueller, E. Roeckl, M.G. Saint-Laurent, and C. Detraz, Z. Phys. A {\bf 327}, 17 (1987)
\bibitem[Pro72]{Pro72t} I.D. Proctor, W. Benenson, J. Dreisbach, E. Kashy, G.F. Trentelman, and B.M. Preedom, Phys. Rev. Lett. {\bf 29}, 434 (1972)
\bibitem[Ric57]{Ric57t} E. Ricci, J.P. Campa, and N. Nussis, Publs. Com. Nacl. Energia Atomica (Buenos Aires) Ser. quim {\bf 1}, 129 (1957)
\bibitem[Roy57]{Roy57t} J.-C. Roy and T.P. Kohman, Can. J. Phys. {\bf 35}, 649 (1957)
\bibitem[Tri77]{Tri77t} R.E. Tribble, J.D. Cossairt, D.P. May,and R.A. Kenefick, Phys. Rev. C {\bf 16}, 917 (1977)
\bibitem[Web92]{Web92t} M. Weber, C. Donzaud, J.P. Dufour, H. Geissel, A. Grewe, D. Guillemaud-Mueller, H. Keller, M. Lewitowicz, A. Magel, A.C. Mueller, G. M\"unzenberg, F. Nickel, M. Pf\"utzner, A. Piechaczek, M. Pravikoff, E. Roeckl, K. Rykaczewski, M.G. Saint-Laurent, I. Schall, C. Stephan, K. S\"ummerer, L. Tassan-Got, D.J. Vieira, and B. Voss, Z. Phys. A {\bf 343}, 67 (1992)

\end{theDTbibliography}

\end{document}